\documentclass[aps,prb,reprint,groupedaddress]{revtex4-2}

\usepackage{graphicx}
\usepackage{bm}

\begin{document}


\title{
Nuclear spin-lattice relaxation rate near a two-dimensional antiferromagnetic quantum critical point
}

\author{Yutaka Itoh}
\affiliation{Department of Physics, Graduate School of Science, Kyoto Sangyo University, Kamigamo-Motoyama, Kita-ku, Kyoto 603-8555, Japan}
 
\date{\today}

\begin{abstract}
We study the effects of zero-point spin fluctuations on the nuclear spin-lattice relaxation rate 1/$T_1$ around a two-dimensional antiferromagnetic quantum critical point in the self-consistent renormalization theory.
For a weak mode-mode coupling constant \mbox{$y_1 < $ 0.1}, the finite temperature behavior of 1/$T_1$ across the quantum critical point resembles that for only the thermal spin fluctuations.  
For a strong mode-mode coupling constant \mbox{$y_1 > $ 0.1}, 1/$T_1$ takes its local maximum as a function of temperature in the nearly antiferromagnetic state. 
Experimental observation of the local maximum in the $^{63}$Cu nuclear spin-lattice relaxation rate 1/$T_1$ of lightly-doped superconductors La$_{2-x}$Sr$_x$CuO$_4$ with $x$ = 0.06--0.10 is associated with the effects of the zero-point spin fluctuations. 
The two-dimensional magnetic quantum critical point and the role of the zero-point spin fluctuations are discussed in the electron spin dynamics of underdoped superconductors La$_{2-x}$Sr$_x$CuO$_4$. 
\end{abstract}


\maketitle

\section{Introduction}
Nuclear spin-lattice relaxation rate 1/$T_1$ due to electron spin fluctuations is expressed in terms of the dynamical spin susceptibility, which characterizes both itinerant electron systems and exchange-coupled local moment systems in a unified manner~\cite{Moriya}.
The conventional Korringa relation of $T_1$ in Fermi liquid does not hold in the metals close to quantum critical points (QCP)~\cite{MU1,MU2}.
The nuclear spin-lattice relaxation rate 1/$T_1$ due to two-dimensional antiferromagnetic spin fluctuations in metals is expressed by the static staggered spin susceptibility $\chi(Q)$ ($Q$ is an antiferromagnetic wave vector) as 1/$T_1$ $\propto T\chi(Q)$ ($T$ is temperature)~\cite{MTU,MMP}. 
The experimental measurements of the nuclear spin-spin relaxation rate 1/$T_{2G}$ confirm the relation of 1/$T_1$ $\propto T\chi(Q)$ in the normal state of layered high-$T_\mathrm{c}$ cuprate superconductors~\cite{YI1,YI2,YI3}.    
Theoretical and experimental efforts have been devoted to understand $\chi(Q)$ in the two-dimensional itinerant electron systems, which are associated with magnetic instability and superconductivity~\cite{SCA}. 

The self-consistent renormalization (SCR) theory is a powerful theory to account for finite temperature itinerant electron magnetism~\cite{MT}.
Although the original electron spin fluctuation spectrum consists of thermal spin fluctuations with Bose factor and zero-point quantum spin fluctuations, 
the traditional SCR theory treats only the thermal spin fluctuations and successfully accounts for finite temperature Curie--Weiss behaviors of static spin susceptibility and
quantum critical behaviors in experimental observations~\cite{MU2,MT}. 
The SCR formalism is justified by renormalization of SCR parameters with the zero-point spin fluctuations in two- and three-dimensional ferromagnetic metals and three-dimensional antiferromagnetic metals~\cite{IM1}.
However, the significant effects of the zero-point spin fluctuations are found in high-field magnetization curves in ferromagnetic systems~\cite{TY0,TY1,TY2}, 
staggered spin susceptibility $\chi(Q)$ and nuclear spin--lattice relaxation rate 1/$T_1$ at a two-dimensional quantum critical point~\cite{IM1}. 
The effect of the zero-point spin fluctuations on $\chi(Q)$ just at the two-dimensional QCP is also found to depend on the strength of a mode-mode coupling constant $y_1$~\cite{YISCR}. 
The value of $y_1$ = 0.1 can be used for identification of the effect.  
Ishigaki and Moriya report that the zero-point spin fluctuations bring about significant change in quantum critical behavior of the nuclear spin-lattice relaxation rate 1/$T_1$ with some typical SCR parameters~\cite{IM1}.

The detailed calculations of asymptotic solutions and numerical solutions of the uniform spin susceptibility $\chi(0)$ and the nuclear spin-lattice relaxation rate 1/$T_1$ 
are reported in the SCR theory around two- and three-dimensional ferromagnetic QCP's and three-dimensional antiferromagnetic QCP~\cite{HM,IM3}.
For nearly ferromagnetic metals, 1/$T_1$ takes its local maximum at a certain temperature near the ferromagnetic QCP~\cite{HM}. 
For three-dimensional nearly antiferromagnetc metals, 1/$T_1$ does not take such a local maximum and decreases $\propto T^{1/4}$ with cooling just at the QCP~\cite{IM3}.       

In this paper, we report on the detailed study of the zero-point spin fluctuation effects on the nuclear spin--lattice relaxation rate 1/$T_1$ due to the two-dimensional antiferromagnetic spin fluctuations near the QCP in the Watanabe--Miyake SCR theory~\cite{WM}. 
The quantum critical behavior of 1/$T_1$  depends on the value of $y_1$. 
For the weak mode-mode coupling constant \mbox{$y_1 < $ 0.1}, 1/$T_1$ near the QCP is nearly the same as that in the absence of the zero-point fluctuations in the Moriya-Takahashi-Ueda SCR theory~\cite{MTU}.      
For the strong mode-mode coupling constant \mbox{$y_1 > $ 0.1}, we find a local maximum in 1/$T_1$ as a function of temperature near the QCP. 
Experimental observation of the local maximum of $^{63}$Cu nuclear spin--lattice relaxation rate 1/$T_1$ in underdoped superconductors La$_{2-x}$Sr$_x$CuO$_4$ with $x$ = 0.06--0.10 is associated with the effects of the zero-point spin fluctuations. 

\section{SCR Theory with Zero-Point Spin Fluctuations near a Two-Dimensional QCP}
\subsection{SCR Equation near a Two-Dimensional Antiferromagnetic QCP}
We set energy units for temperature $T$ and frequency $\omega$ by $k_{\mathrm B}$ and ${\hbar}$ = 1, and
$\chi(Q)$ in units of $(2\mu_\mathrm{B})^2$.
For convenience, the notations of the spin fluctuation parameters conform with the Watanabe--Miyake SCR theory~\cite{WM}.  

The dynamical spin susceptibility $\chi(Q+q, \omega)$ peaked at $Q$ is expanded as 
\begin{equation}
\chi(Q+q, \omega) \approx \frac{1/2T_A(y + x^2)}{1 - i\omega/2\pi T_0(y + x^2)},
\label{Xqw}
\end{equation}
where $x$ = $q/q_B$ ($q_B$ is the effective Brillouin zone boundary) is the~reduced wave number, $y$ = 1/2$T_A\chi(Q)$ = 1/$(\xi q_B)^2$ ($\xi$ is the antiferromagnetic correlation length) is the~reduced inverse staggered spin susceptibility, $T_A$ is the momentum spread of the spin fluctuation spectrum, and $T_0$ is the frequency spread of the spin fluctuation spectrum.
The amplitude of the spin fluctuation spectrum is $\chi(Q) = 1/2\pi T_Ay$. 
The spin fluctuation frequency is $\Gamma(Q) = 2\pi T_0 y$.

The Watanabe--Miyake SCR equation of $y$ against the reduced temperature $t$ = $T/T_0$ near the two-dimensional antiferromagnetic QCP is given by  
\begin{equation}
y = y_0 + \frac{y_1}{2}\Big[ (y{\rm ln}y - c y) +\frac{t}{2}\Big\{{\rm ln}\frac{x_c^2 + y}{y}-{\rm ln}\frac{x_c^2 + y + t/6}{y + t/6}\Big\}\Big],
\label{eq1} 
\end{equation} 
where $x_c$ = $q_c/q_B$ is the cut-off wave number~\cite{WM,IM1} and $c$ = 1 + 2${\rm ln}x_c$~\cite{IM1}. 
In the Watanabe-Miyake SCR theory, $y_0$ is the ground state parameter defined as the zero-temperature inverse staggered spin susceptibility, in which logarithmic correction is not renormalized~\cite{WM}.
$y_0$ is a measure of the distance to the QCP $y_0$ = 0. 
$y_0 >$ 0 corresponds to the paramagnetic ground state, while $y_0 <$ 0 corresponds to the antiferromagnetic ground state~\cite{MTU}.
In the Ishigaki-Moriya SCR theory, $y(t = 0)$ is denoted by $y_0$ in their two-dimensional SCR equation~\cite{IM1}. 
$y_1$ is a mode-mode coupling constant for the staggered modes with $Q$. 
The reduced inverse staggered spin susceptibility $y$ is self-consistently determined by Equation~(\ref{eq1}) as a function of $t$ for given $y_0$ and $y_1$. 
The SCR parameters are $y_0$, $y_1$, $T_A$, and $T_0$. 

In Equation~\ref{eq1}, $(y{\rm ln}y - c y)$ and $(t/2)\{\cdots\}$ result from the two-dimensional zero-point spin fluctuations and the thermal fluctuations, respectively.
The term of $(y{\rm ln}y - c y)$ can cause significant effects on $\chi(Q)$ and 1/$T_1$ near the two-dimensional QCP~\cite{IM1}. 

The nuclear spin-lattice relaxation rate 1/$T_1$ due to the two-dimensional antiferromagnetic spin fluctuations is expressed as
\begin{equation}
\frac{1}{T_1} = \frac{h}{k_B}\Big(\frac{\gamma_N}{2\pi}\Big)^2\frac{A_{hf}^2}{T_A}\frac{1}{\overline{T}_1}.
\label{T1}
\end{equation} 
where $\gamma_N$ is the nuclear gyromagnetic ratio, $A_{hf}$ is the hyperfine coupling constant, and 1/$\overline{T}_1 \equiv t/y$ is the reduced nuclear spin-lattice relaxation rate~\cite{MTU,IM1}.
We calculate and discuss the reduced nuclear spin-lattice relaxation rate 1/$\overline{T}_1$ near the two-dimensional QCP.

\subsection{Zero--Temperature SCR Equation}
The~SCR Equation~(\ref{eq1}) indicates that $y$(0) defined by $y(t = 0)$ at zero temperature disagrees with the input value of $y_0$.
The SCR Equation~(\ref{eq1}) for $y(0)$ leads to a zero-temperature SCR equation of
\begin{equation}
y(0) = y_0 + \frac{y_1}{2}\Big[ y(0){\rm ln}y(0) - c y(0)\Big],
\label{yt0} 
\end{equation}
where $y$(0) is self-consistently determined for the input values of $y_1$ and $y_0$.
Alternatively, the zero--temperature SCR Equation~(\ref{yt0}) can be rewritten as 
\begin{equation}
 y_0 = y(0) - \frac{y_1}{2}\Big[ y(0){\rm ln}y(0) - c y(0))\Big],
\label{y0} 
\end{equation}
which is an expression of $y_0$ as a function of $y$(0).
\begin{figure}[h]
 \begin{center}
\includegraphics[width=7.0 cm]{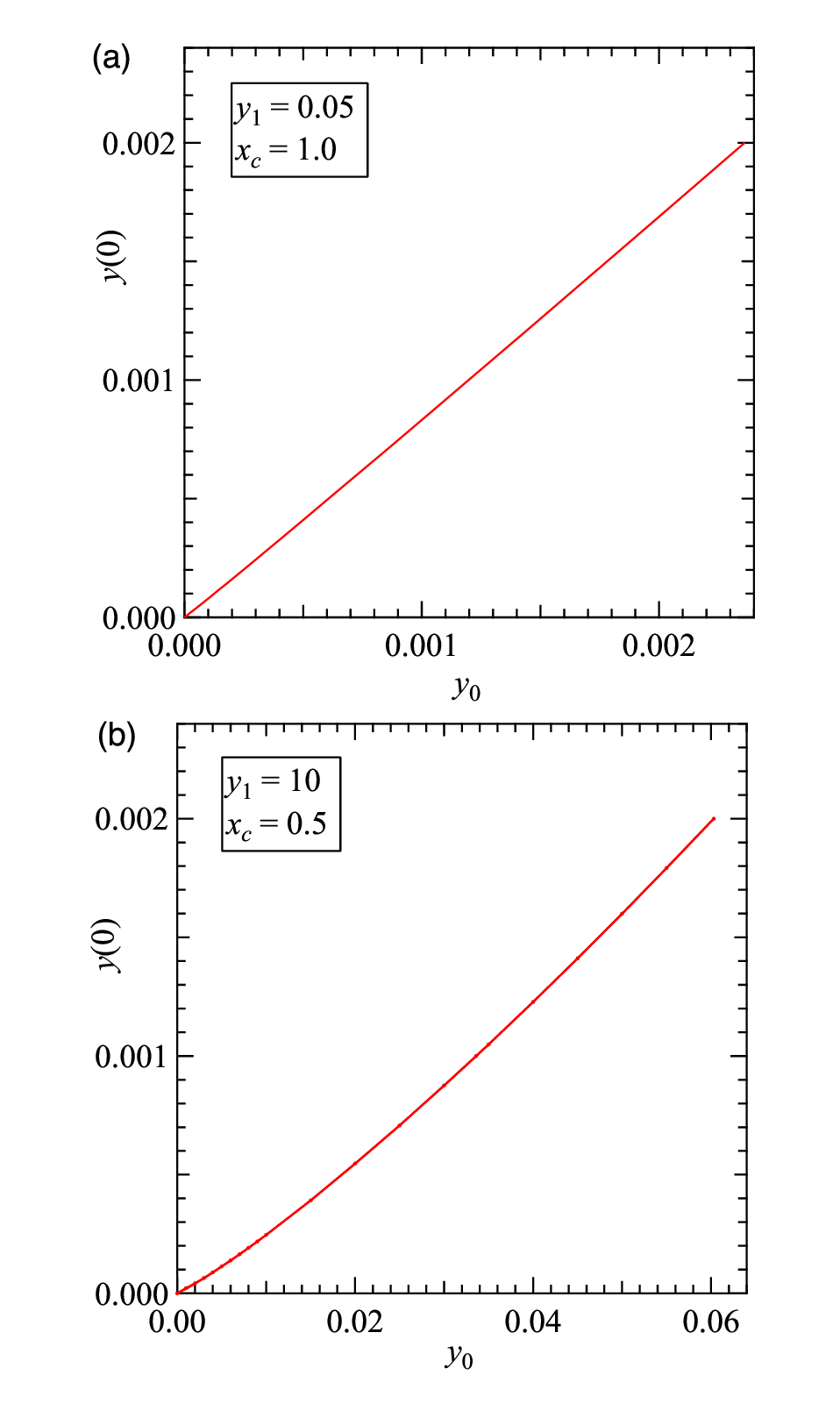}
 \end{center}
\caption{
$y(0)$ vs. $y_0$ for $y_1$ = 0.05 and $x_c$ = 1.0 (\textbf{a}) and $y_1$ = 10 and $x_c$ = 0.5 (\textbf{b}).  
\label{y0y0}
}
\end{figure}

Figure~\ref{y0y0} shows the numerical calculations of $y(0)$ vs. $y_0$ for $y_1$ = 0.05 and $x_c$ = 1.0 (\textbf{a}) and $y_1$ = 10 and $x_c$ = 0.5 (\textbf{b}).   
One should note $y_0 >$ $y$(0) due to the logarithmic term of the zero-point spin fluctuations. 
In the two- and three-dimensional ferromagnetic spin fluctuations and the three-dimensional antiferromagnetic spin fluctuations,
the ground state parameter $y_0 = y(0)$ holds at zero temperature~\cite{IM1}.  

\subsection{Asymptotic Solutions of SCR Equation near a Two-Dimensional Antiferromagnetic QCP}
The SCR Equation~(\ref{eq1}) near the two-dimensional antiferromagnetic QCP has asymptotic solutions, as in the case of the ferromagnetic QCP~\cite{HM}.

For $y_0 <$ 0, since the ground state is in antiferromagnetic long range ordering with the Neel temperature $T_N$ = 0 K,
the asymptotic solution of the SCR Equation~(\ref{eq1}) is 
\begin{equation}
y(t) \sim t~\mathrm{exp}(- 4|y_0|/y_1t)/6,
\label{RC} 
\end{equation}
which indicates critical slowing down of the spin fluctuation frequency $\Gamma(Q) = 2\pi T_0 y$~\cite{MTU,TM}.
The Equation~(\ref{RC}) is derived from balance between the negative $y_0$ and the thermal fluctuation terms, where
the zero-point fluctuation term $(y{\rm ln}y - c y)$ is negligible. 

For $y_0$ = 0, the zero-temperature quantum critical behavior is the asymptotic solution of $y$($t$) = $-t$ln$|$ln$t|$/2ln$t$~\cite{IM1,WM}.
The finite temperature asymptotic solution is an inverse Curie law in the weak mode--mode coupling limit $y_1 \rightarrow$ 0 as 
\begin{equation}
y(t) \sim (-y_1\mathrm{ln}y_1)t/4,
\label{QC} 
\end{equation}
where the zero--point fluctuation term $(y{\rm ln}y - c y)$ is also negligible~\cite{YISCR}. 

For $y_0 >$ 0, since the ground state is in a nearly antiferromagnetic paramagnetic state, at $t \rightarrow$ 0 and $t \ll y(0)$, we obtain the asymptotic solution for any $y_1$ as
\begin{eqnarray}
y(t) &\sim y(0)[1 + (t/t^{\ast})^2], \\
t^{\ast} &= y_0\sqrt{\frac{24}{y_1}},
\label{FL} 
\end{eqnarray}
where $t^{\ast}$ is a crossover temperature toward low temperature Fermi liquid~\cite{Ueda}. 
One should note that $y(0)$ is a renormalized value with logarithmic correction of the zero--point spin fluctuations. 
Then, as to the low temperature behavior $y(t < t^{\ast})$ for $y_0 >$ 0, the expression~(\ref{FL}) for the zero-point spin fluctuations is the same form as that for only the thermal spin fluctuations~\cite{MTU,Ueda}. 
Below $t^{\ast}$, $y(t)$ goes to $y(0)$.  
Since $y(t^{\ast})$ = 2$y(0)$, $t^{\ast}$ also indicates deviation from the high temperature Curie-Weiss law in $\chi(Q)$.  

As $y_0$ approaches 0, neither the asymptotic solution for $y_0 <$ 0 nor that for $y_0 >$ 0 converges to the logarithmic solution at $y_0$ = 0.
The QCP asymptotic solution of $y$($t$) = $-t$ln$|$ln$t$$|$/2ln$t$ is an analytically disconnected function.
Thus, we study numerically the finite temperature SCR Equation~(\ref{eq1}) in the vicinity of the QCP. 

\section{Numerical Inverse Staggered Spin Susceptibility \boldmath$y$ and Nuclear Spin-Lattice Relaxation Rate 1/$T_1$ from Weak to Strong Mode-Mode Coupling Constant \boldmath$y_1$ }
We solved numerically the SCR Equation~(\ref{eq1}) of $y$ as a function of $t$ with given values of $y_0$ 
for weak and strong mode-mode coupling constants of $y_1$ = 0.05 ($<$ 0.1) and 10 ($>$ 0.1).
Here, $y_1$ = 0.1 is a criterion to classify the zero-point fluctuation effects~\cite{YISCR}. 
And then, we calculated the reduced nuclear spin--lattice relaxation rate 1/$\overline{T}_1 \equiv t/y$ with the numerical $y$ and $t$ from weak to strong mode-mode coupling regimes. 

\subsection{Weak Mode-Mode Coupling}
\begin{figure}[t]
 \begin{center}
\includegraphics[width=8.0 cm]{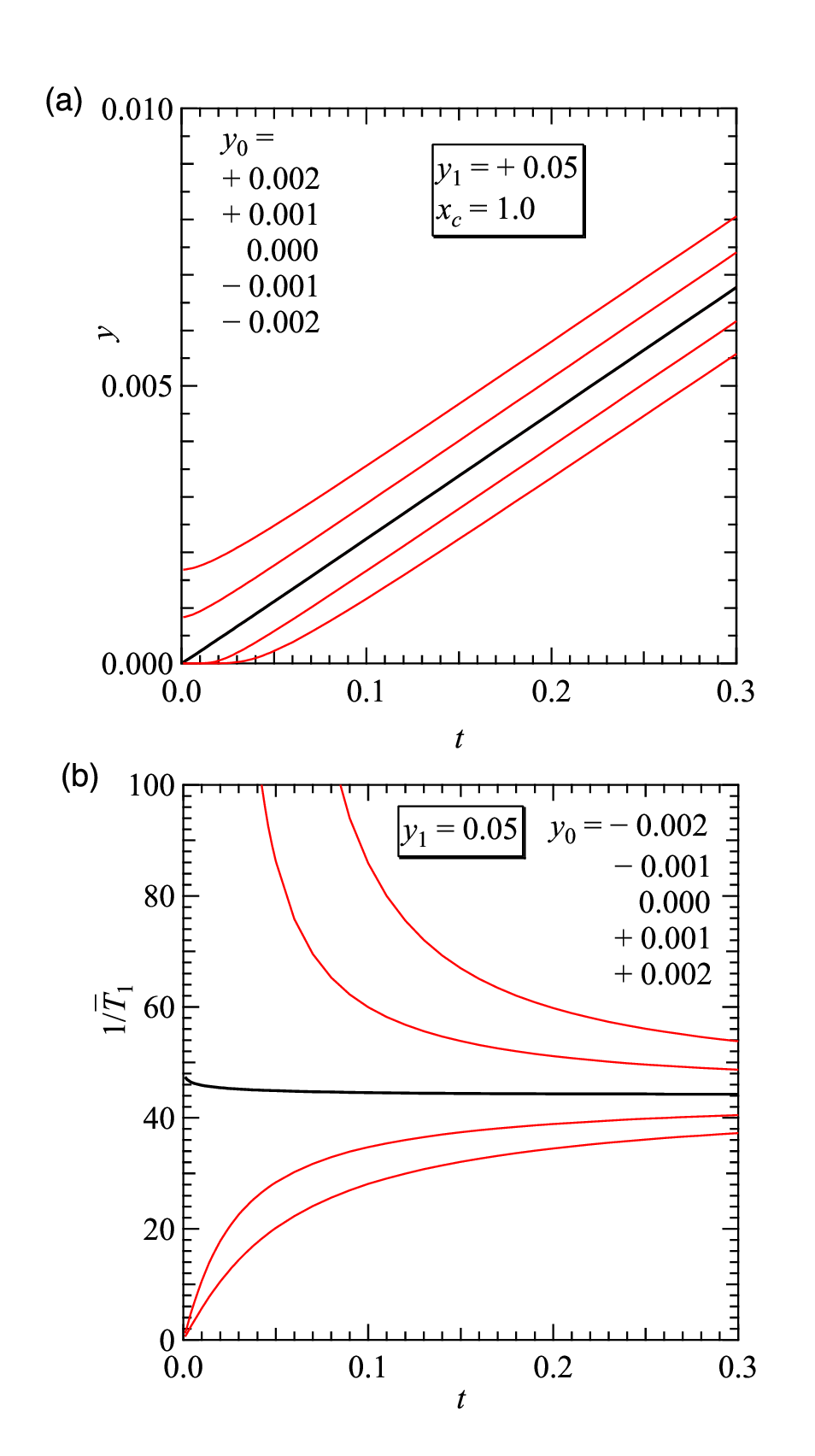}
\vspace*{+5.0mm}
 \end{center}
\caption{
Weak mode--mode coupling $y_1$ = 0.05 in the range of $y_0$ from $-$0.002 to $+$0.002 : 
(\textbf{a}) 
$y$ plotted against $t$.
(\textbf{b}) 
1/$\overline{T}_1$ plotted against $t$.  
\label{WC}
}
\end{figure}

Figure~\ref{WC}a shows the reduced inverse staggered spin susceptibility $y$ against the reduced temperature $t$ for  $y_0$ from $-$0.002 to $+$0.002 in increments of 0.001 and $y_1$ = 0.05. 
Figure~\ref{WC}b shows the reduced nuclear spin--lattice relaxation rate 1/$\overline{T}_1$ against $t$ for $y_0$ from $-$0.002 to $+$0.002 and $y_1$ = 0.05. 

The present weak coupling results of $y$ and 1/$\overline{T}_1$ as functions of $t$ for various $y_0$ are parallel to those in the Moriya--Takahashi--Ueda SCR theory~\cite{MTU}.
The high-temperature 1/$\overline{T}_1$ is nearly a temperature-independent constant for each $y_0$.
However, there are some differences from the SCR results with only the thermal fluctuations. 
In Figure~\ref{WC}b, just at the QCP $y_0$ = 0, the low-temperature 1/$\overline{T}_1$ shows a slight increase.
The upturn at $t \rightarrow$ 0 is the effect of the zero-point spin fluctuations.   
The quantum critical behavior of 1/$\overline{T}_1$ across the QCP is more sensitive to the value of $y_0$ than it is in the case of thermal fluctuations alone.   

\subsection{Strong Mode-Mode Coupling}
\begin{figure}[t]
 \begin{center}
\includegraphics[width=8.0 cm]{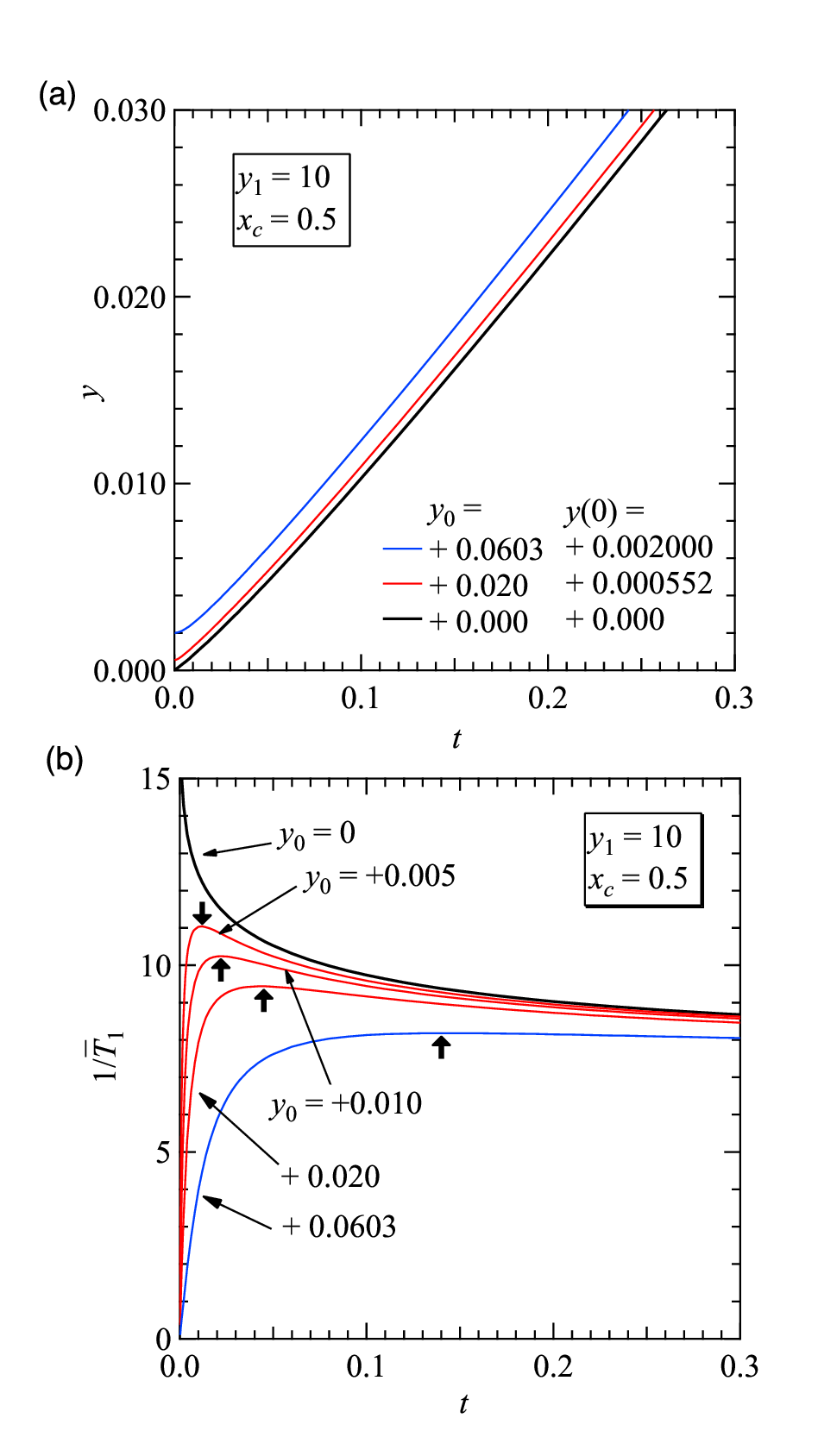}
 \end{center}
\caption{
Strong mode--mode coupling $y_1$ = 10 in the range of $y_0$ from 0 to $+$0.0603 : 
(\textbf{a}) 
$y$ plotted against $t$. 
(\textbf{b}) 
1/$\overline{T}_1$ plotted against $t$. 
Short arrows indicate the temperature $t_{max}$ at which 1/$\overline{T}_1$ takes its local maximum. 
\label{SC}
}
\end{figure}

Figure~\ref{SC}a shows the reduced inverse staggered spin susceptibility $y$ against the reduced temperature $t$ for strong coupling $y_1$ = 10.
The present results of $y$ vs. $t$ are in the nearly antiferromganetic state of 0 $\leq$ $y_0$ $\leq$ $+$0.0603. 
Figure~\ref{SC}b shows the reduced nuclear spin--lattice relaxation rate 1/$\overline{T}_1$ against $t$.
The results for 1/$\overline{T}_1$ obtained in this study, where $y$(0) (the corresponding $y_0$) is 0.000128 (0.005), 0.000256 (0.010), and 0.000552 (0.020), fill in the gap in the data between $y$(0) = 0 and 0.001 in the previous Ishigaki-Moriya study~\cite{IM1}.    

In Figure~\ref{SC}b, just at the QCP $y_0$ = 0, 1/$\overline{T}_1$ shows logarithmic divergence toward zero temperature~\cite{IM1}.
For $y_0 >$ 0, 1/$\overline{T}_1$ shows its local maximum at a certain temperature, which we denote by $t_{max}$ (short arrows).
As $y_0$ increases from 0 to +0.0603, a narrow maximum of 1/$\overline{T}_1$ changes to a broad maximum, 
and $t_{max}$ increaes.
Since the strong mode--mode coupling enhances the effects of the zero--point fluctuations,
the maximum behavior in 1/$\overline{T}_1$ is also attributed to development of the zero--point spin fluctuations~\cite{YISCR}.  

\begin{figure}[h]
\includegraphics[width=7.0 cm]{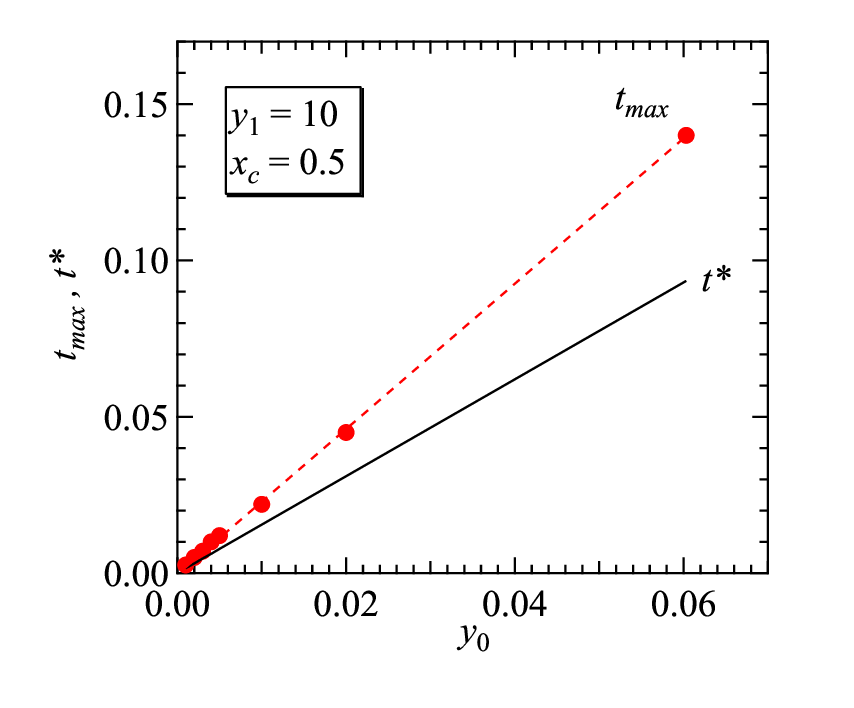}
\caption{
$t_{max}$ (closed circles) and crossover temperature $t^{\ast}$ against $y_0$.
A broken line is a linear fit function to $t_{max}$. 
A solid line is $t^{\ast}$ as a function of $y_0$ with $y_1$ = 10.   
\label{fig4}
}
\end{figure}

Figure~\ref{fig4} shows $t_{max}$ against $y_0$ and the Fermi-liquid crossover temperature $t^{\ast}$ of Equation~(\ref{FL}) against $y_0$ with $y_1$ = 10.
$t_{max}$ increases linearly with increasing $y_0$, as shown by a broken line.

\vspace{-3pt}
\section{Zero-Point Spin Fluctuations in La$_{2-x}$Sr$_{x}$CuO$_4$} 

La$_{2-x}$Sr$_x$CuO$_4$ is one of the most intensively studied high-$T_\mathrm{c}$ superconductors. 
$x$ is the Sr content and the number of the doped holes.
The electrical resistivity measurements indicate that $x_\mathrm{SI}$ = 0.04 is the superconductor-to-insulator boundary~\cite{Takagi}. 
Although the nuclear magnetic relaxation measurements have revealed the two-dimensional antiferromagnetic quantum critical behaviors and advanced the spin fluctuation theory~\cite{Imai,Ohsugi,Imai3}, 
the reason why 1/$T_1$ exhibits a local maximum at a certain temperature between $T_\mathrm{c}$ and the high-temperature region in low-doping superconductors remained unclear.
\vspace{-3pt}
\begin{figure}[t]
\hspace*{+10 mm}
\includegraphics[width=8.5 cm]{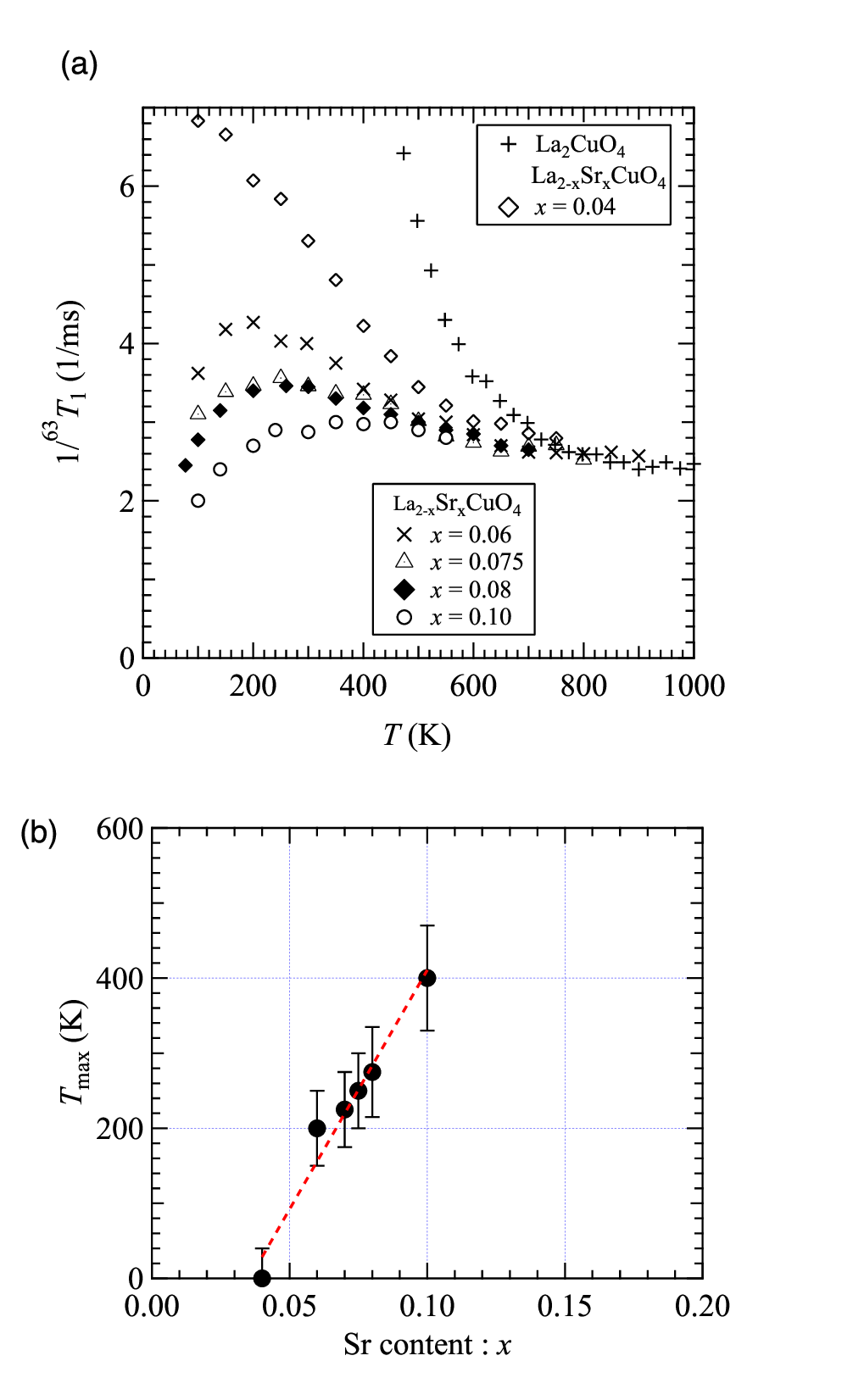}
\caption{
(\textbf{a}) Experimental $^{63}$Cu nuclear spin--lattice relaxation rate 1/$^{63}T_1$ plotted against $T$ for lightly--doped La$_{2-x}$Sr$_{x}$CuO$_{4}$, reproduced from the experimental reults~\cite{Imai,Fuji,Itoh2,Baek}.
For $x$ = 0.06--0.10, 1/$^{63}T_1$ shows its local maximum and approaches a constant value at high temperatures. 
(\textbf{b}) Closed circles are $T_{max}$ at which 1/$^{63}T_1$ takes its local maximum. 
A broken line is the best fit function of $T_{max}/$K = 6371($x$ $-$ 0.035).  
\label{LSCOT1}
}
\end{figure}

Figure~\ref{LSCOT1}a shows the experimental $^{63}$Cu nuclear spin--lattice relaxation rate 1/$^{63}T_1$ for lightly--doped La$_{2-x}$Sr$_{x}$CuO$_{4}$ reproduced from the experimental results~\cite{Imai,Fuji,Itoh2,Baek}. 
1/$^{63}T_1$ takes its local maximum at $T_{max}$ = 200 K for $x$ = 0.06. 
As $x$ increases, $T_{max}$ rises to about 400 K for $x$ = 0.10.
Such a local maximum is also found in 1/$\overline{T}_1 \equiv T(1/T_G)^2$ with Gaussian decay rate 1/$T_G$ for underdoped YBa$_2$Cu$_3$O$_{6.5}$ and YBa$_2$Cu$_4$O$_8$~\cite{Itoh}.
The local maximum in 1/$^{63}T_1$ emerges at higher temperatures than the spin pseudo-gap and the slow spin dynamics at low temperatures~\cite{ItohLSCO}.    

Figure~\ref{LSCOT1}b shows $T_{max}$ vs. $x$. 
The broken line is the best fit function of $T_{max}/$K = 6317($x$ $-$ 0.035), which indicates the critical Sr content $x_{c}$ = 0.035 as the QCP $y_0$ = 0.
The estimated value of $x_{c}$ = 0.035 is close to the superconductor-to-insulator boundary $x_\mathrm{SI}$ = 0.04 and the onset of spin density wave order~\cite{SD}. 

Precisely, we could not reproduce detailed temperature dependence of 1/$^{63}T_1$ from the numerical calculations of 1/$\overline{T}_1$,
although some 1/$^{63}T_1$ component in the frequency distribution~\cite{Imai3} might fit to the theoretical curves in Figure~\ref{SC}b. 
It is worth noting that the behavior of the local maximum at $T_{max}$ and the constant behavior in 1/$^{63}T_1$ at high temperatures are similar to the behavior in 1/$\overline{T}_1$ near the QCP shown in Figure~\ref{SC}a.
From application of a typical $t_{max} = T_{max}/T_0$ = 0.10 to $T_{max}$ = 200--300 K, the spin fluctuation energy $T_0$ is estimated as 2000--3000 K.
These are sound values. 
These results suggest that the zero--point spin fluctuations have the significant effects on the electron spin dynamics in La$_{2-x}$Sr$_{x}$CuO$_{4}$.

Antiferromagnetic spin fluctuations can have both pairing and depairing effects on superconductivity~\cite{Ohashi,Ueda,MU1,MU2}. 
The zero-point spin fluctuations may enhance low frequency components of the imaginary part of the dynamical spin susceptibility $\chi(q, \omega)$. 
The low frequency spin fluctuations can cause depairing effect on superconductivity.
The high frequency spin fluctuation leads to the high $T_\mathrm{c}$.
The reason why La$_{2-x}$Sr$_{x}$CuO$_{4}$ with a large spin fluctuation energy $T_0$ has relatively lower $T_\mathrm{c}$ than HgBa$_2$CuO$_{6+\delta}$ may be attributed to depairing effect of the zero-point spin fluctuations. 

\section{Conclusions}
In conclusion, the two-dimensional antiferromagnetic quantum critical behavior of the reduced nuclear spin-lattice relaxation rate 1/$\overline{T}_1$ depends on the value of the mode--mode coupling constant $y_1$ through the zero-point spin fluctuations. 
The quantum critical behavior of 1/$\overline{T}_1$  is sensitive to $y_1$. 
In the weakly coupling regime (0 $< y_1 <$ 0.1), 1/$\overline{T}_1$ is nearly the same as that in only the thermal fluctuations.  
In the strongly coupling regime ($y_1 >$ 0.1), due to the zero-point spin fluctuations, 1/$\overline{T}_1$ takes its local maximum between low and high temperature regions.
Such a local maximum in 1/$\overline{T}_1$ in the vicinity of the QCP is associated with the experimental observations in La$_{2-x}$Sr$_{x}$CuO$_{4}$ with $x$ = 0.06--0.10.




\end{document}